\pdfoutput=1
\documentclass{ws-jnopm}

\usepackage{graphicx}
\graphicspath{{pict/}{}}

\newcommand{\be}{\begin{equation}}
\newcommand{\ee}{\end{equation}}

\newcommand\pictc[5]{\begin{figure}[t,b]
                   \centerline{
\includegraphics[width=#1\columnwidth,height=0.7\textheight,keepaspectratio]{#3}}
               \protect\caption{\protect\label{fig:#4} #5}
                \end{figure}            }

\newcommand\pict[4][0.99]{\pictc{#1}{!tb}{#2}{#3}{#4}}
\newcommand\rpict[1]{\ref{fig:#1}}

\newcommand\leqt[1]{\protect\label{eq:#1}}
\newcommand\reqtn[1]{\ref{eq:#1}}
\newcommand\reqt[1]{(\reqtn{#1})}

\title{Nonlinear optics and light localization in periodic photonic lattices}

\author{Dragomir N. Neshev$^1$, Andrey A. Sukhorukov$^1$, Wieslaw Krolikowski$^2$, and Yuri S. Kivshar$^1$}
\address{$^1$Nonlinear Physics Centre, Research School of Physical
Sciences and Engineering, \\Centre for Ultra-high bandwidth Devices
for Optical Systems (CUDOS),  \\Australian National University,
Canberra ACT 0200, Australia\\
$^2$Laser Physics Centre, Research
School of Physical Sciences and Engineering, \\Centre for Ultra-high
bandwidth Devices for Optical Systems (CUDOS),
\\Australian National University, Canberra ACT 0200, Australia}

\begin{document}

\maketitle

\begin{history}
\end{history}

\begin{abstract}
We review the recent developments in the field of photonic lattices emphasizing their unique properties for controlling linear and nonlinear propagation of light. We draw some important links between optical lattices and photonic crystals pointing towards practical applications in optical communications and computing, beam shaping, and bio-sensing.
\end{abstract}

\keywords{Optical lattices, bandgap, nonlinear self-action, optical solitons}

\markboth{D. N. Neshev et al.}{Nonlinear Optics and Light Localization in Periodic Photonic Lattices}

\section{Optical lattices as nonlinear photonic crystals}
\label{sect:intro}

Nonlinear propagation of light in periodic structures has become an attractive area of research in recent years\cite{Kivshar:2003:OpticalSolitons,Slusher:2003:NonlinearPhotonic,Soljacic:2004-211:NAMT}, holding strong promises for novel applications in photonics. The underlying physical effects are analogous to those occurring in a number of different systems, including biological molecular structures, solid-state systems, and Bose-Einstein condensate in periodic potentials. A special class of periodic structures are {\em photonic crystals}\cite{Joannopoulos:1995:PhotonicCrystals}, artificial materials with optical bandgaps, which offer unique possibilities for controlling the propagation of light in a way similar to that of semiconductors used for manipulating the flow of electrons. Photonic crystals represent a broad class of structures with periodicity of the refractive index on the wavelength scale in one, two, or three dimensions. They were first suggested in the pioneering papers of Eli Yablonovitch\cite{Yablonovitch:1987-2059:PRL} and Sajeev John\cite{John:1987-2486:PRL}, and nowadays this terminology is widely applicable to many different materials (see Fig.~\rpict{photonic_crystals}), some of which are commonly used in almost any optical laboratory. Examples of one-dimensional photonic crystals include dielectric mirrors, Bragg gratings, and arrays of optical waveguides\cite{Slusher:2003:NonlinearPhotonic,Yeh:1988:OpticalWaves}. Two-dimensional photonic crystals are commonly represented by photonic crystal fibers\cite{Russell:2003-358:SCI} and planar photonic crystals\cite{Krauss:1996-699:NAT}. In three dimensions, photonic crystals are known to exist in opal, wood-pile, or inverse opal geometries\cite{Joannopoulos:1995:PhotonicCrystals}. The common feature of all these different structures is that they allow for manipulation of the flow of light in the direction of periodicity. Therefore, photonic crystals offer a possibility to achieve ultimate control over linear and nonlinear properties of the light propagation, as well as enhanced control over light emission and amplification. Herewith we concentrate primarily on the abilities of the periodic structures to control both linear and nonlinear propagation of light.

\pict[0.8]{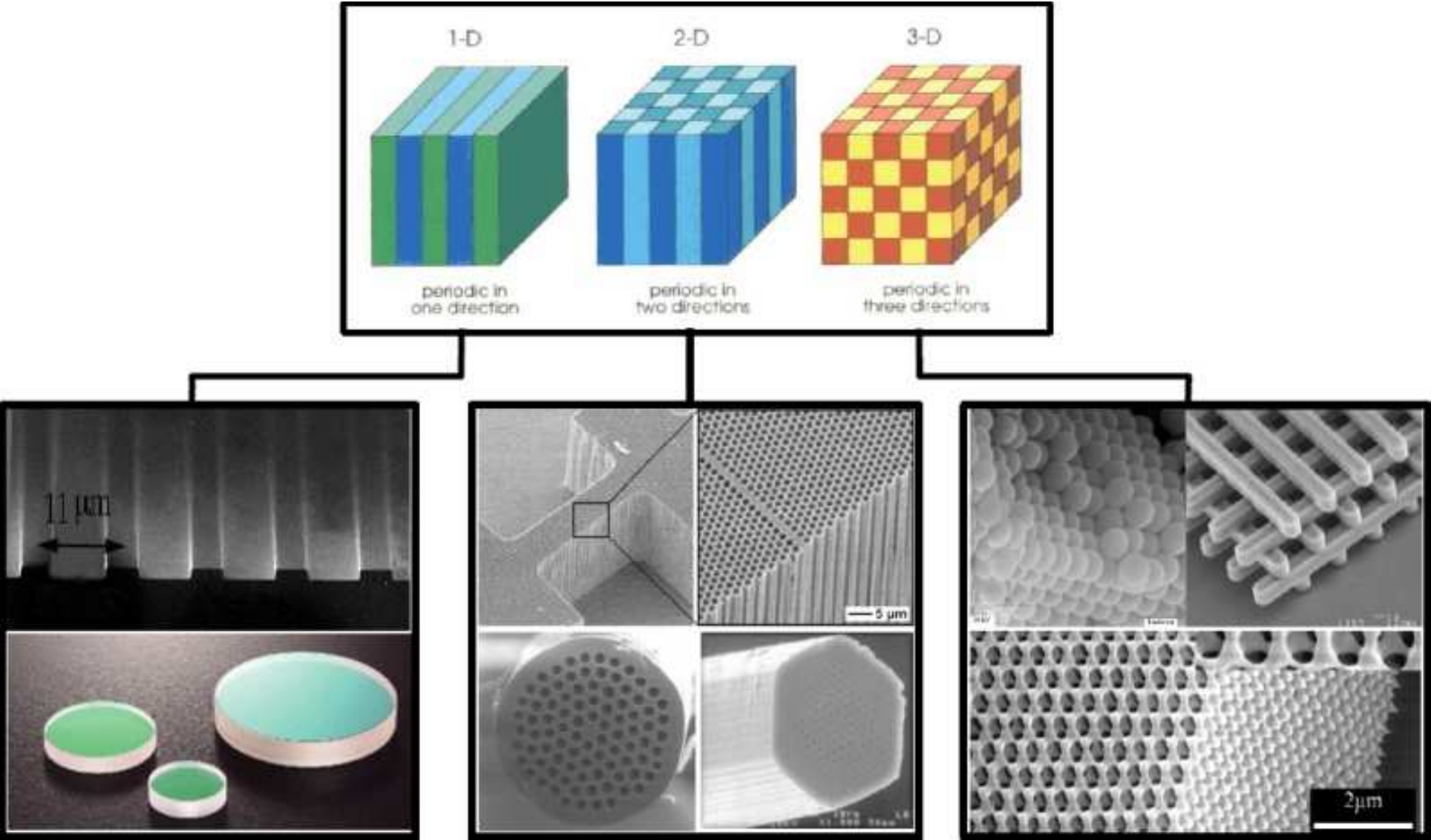}{photonic_crystals}{Examples of one-, two-, and three-dimensional photonic crystals$^{1-4}$.}

Several approaches for controlling the light propagation by engineered periodic structures have been suggested  theoretically and demonstrated experimentally in recent years. These include manipulation of linear light propagation (refraction, diffraction, and dispersion) as well as various nonlinear effects such as harmonic generation, stimulated scattering, and nonlinear self-action. Important examples for a control of refraction and diffraction of light constitute the effects of negative refraction\cite{Cubukcu:2003-604:NAT} and self-collimation\cite{Kosaka:1999-1212:APL}. Furthermore, a design of the structural dispersion allows one to manipulate the group velocity of light resulting in the slow-light propagation\cite{Gersen:2005-73903:PRL}, or enhancement of the spectral response with a fascinating example of the superprism effect\cite{Kosaka:1999-2032:JLT}. A nonlinear response of the material offers novel opportunities for dynamic tunability of the structures by varying the light intensity\cite{Mingaleev:2002-48:OPN}. An interplay between nonlinearity and periodicity represents unique way to efficiently manipulate light by light for optical switching and signal processing applications. Furthermore, light control can be scaled down to micron-scale structures suitable for integration of multiple functionality on a photonic chip.

The goal of our studies in this field is twofold. First, we analyze the fundamentals of nonlinear wave physics in periodic media and, second, through relevant experimental observations we exploit these effects for practical applications. Currently, the major research on photonic structures is concentrated on the demonstration of fundamental physical phenomena. However, novel photonic applications are showing up on the horizon promising a significant impact on various technologies in optical communications and computing, beam steering and shaping, bio-sensing and medical diagnostics.

To achieve our goals we need to employ periodic structures of different geometries, which are easy to fabricate, and which possess strong nonlinearities at moderate laser powers. Currently there exists a number of different approaches for the fabrication of photonic crystals including focused ion beam milling, e-beam lithography combined with reactive ion etching, two-photon and UV polymerization. All these techniques, however, are resource demanding and cost ineffective, imposing great constrains on the fundamental research, where flexible and quick modification of the structural parameters is required. A simpler fabrication and characterization of the periodic photonic structures can be achieved when the scale of periodicity is larger than the wavelength of light. Termed as {\em optical lattices} such periodic photonic structures include optical waveguide arrays\cite{Yariv:1991:OpticalElectronics}, photonic crystal fibers\cite{Russell:2003-358:SCI}, and optically-induced lattices\cite{Efremidis:2002-46602:PRE,Christodoulides:2003-817:NAT,Fleischer:2005-1780:OE}. The scale of periodicity in such structures is of the order of a few micrometers, and their fabrication is facilitated by standard lithographic techniques, fiber drawing, or multiple beam interference. The challenge is to achieve strong nonlinear response of the material at moderate laser powers. The nonlinear response can be enhanced through extended propagation length (as in the case of fibers), stronger light confinement (as in nanowires and photonic crystals), or through slow and resonant nonlinearities (as in the case of optically-induced lattices). In our current studies we chose materials with slow nonlinear response, but with access to nonlinear effects at micro-Watt laser powers. The periodic photonic structures with strong nonlinear response and various geometries allow us to study the basics of nonlinear physics in periodic structures. The obtained knowledge can then be applied to the fabricated photonic structures and photonic crystals in particular, thus bringing the advantages of miniaturization and integration on a photonic chip. We believe that the process of basic knowledge accumulation and its further application is the successful path towards future photonic technologies.

In this paper, we review some recent advances and fundamental concepts of nonlinear light propagation in periodic photonic structures, emphasizing their abilities to control the spatial dynamics of light propagation. The paper is organized as follows. In Section~\ref{sect:linear}, we present the basic concepts of the optically-induced lattices and discuss how their tunability can be used to engineer the linear light propagation such as refraction and diffraction. In Section~\ref{sect:nonlinear}, we review our results on the nonlinear localization of light in periodic structures and outline some novel ideas for the nonlinear control of wave transport in periodic structures. Section~\ref{sect:fabricated} is focused on novel physical phenomena of light propagation in fabricated periodic photonic structures, including the beam interactions with the interfaces.

\section{Fundamentals of light propagation in periodic structures}
\label{sect:linear}

The propagation of light in periodic optical lattices is determined by three major effects: (i) the coupling between neighboring sites of the lattice (inter-site coupling), (ii) Bragg scattering arising from the periodicity of the lattice, and (iii) the specific lattice geometry, being one- (1D), two- (2D), or three-dimensional (3D). The geometry is of particular importance in 2D and 3D, where the lattice symmetry (e.g. square, hexagonal) significantly affects both the inter-site coupling and wave scattering in different directions.
Being able to engineer all those characteristics, one can fully control the propagation of light. Additionally, dynamic tunability of the lattice depth and periodicity in real time will provide an ideal system for experimental studies of fundamental physical phenomena.

\subsection{Optically-induced one- and two-dimensional lattices}

A great opportunity for controlling the lattice parameters was offered by the idea of optically-induced lattices in a biased photorefractive crystal proposed by Efremidis and co-workers\cite{Efremidis:2002-46602:PRE}. With this theoretical proposal, the optical lattice is induced by the interference of two or more broad laser beams propagating inside the photorefractive crystal. The authors took advantage of the strong electro-optic anisotropy of the crystal, such that the ordinarily polarized lattice-forming beams are not affected by the applied electric field. The strong but anisotropic electro-optic effect of the crystal leads to index changes predominantly for the extraordinarily polarized light. Thus, the periodic light pattern resulting from the multiple beam interference will induce a periodic optical potential for any extraordinarily polarized probe beam. At the same time, probe beams will experience strong nonlinear self-action at moderate ($\mu$W) laser powers. The propagation of light in such a system can be described (in isotropic approximation) by the nonlinear Schr\"odinger equation for the slowly varying amplitude of the electric field\cite{Efremidis:2002-46602:PRE,Fleischer:2003-147:NAT,Neshev:2003-710:OL}
\begin{equation} \leqt{nls}
   i \frac{\partial E}{\partial z}
   + D \left( \frac{\partial^2 E}{\partial x^2}
               +\frac{\partial^2 E}{\partial y^2} \right)
    - \frac{\gamma V_0 }{I_b + I_p(x,y) + |E|^2} E = 0,
\end{equation}
where $(x,y)$ and $z$ are the transverse and propagation coordinates normalized to the characteristic values $x_s=y_s= 1\,\mu$m and $z_s = 1$\,mm, respectively, $D = z_s \lambda/(4 \pi n_0 x_s^2)$ is the diffraction coefficient, $\lambda$ is the wavelength in vacuum, $n_0$ is the average refractive index of the medium, and $V_0$ is the bias voltage applied across the crystal. The term  $\gamma V_0 (I_b + I_p(x,y) + |E|^2)^{-1}$ characterizes the total refractive index modulation induced by the optical lattice and the probe beam, linearly proportional to the applied electric field. Here $I_b=1$ is the normalized constant dark irradiance, and the lattice intensity $I_p$ depends on the specific lattice geometry controlled by the number and position of the lattice forming beams. For example, $I_p(x,y)=I_g |\exp(i k x) + \exp(- i k x )|^2$ for 1D lattice produced by two-beam interference [Fig.~\rpict{lattices}(a)]; $I_p(x,y)=I_g |\exp(i k x)
            + \exp(- i k x )
            + \exp(i k y)
            + \exp(- i k y)|^2$
for a 2D lattice of square geometry created by four coherent beams [Fig.~\rpict{lattices}(b)]; or $I_p(x,y)=I_g |\exp(i k x)
            + \exp(- i k x / 2 - i k y \sqrt{3}/2  )
            + \exp(- i k x / 2 + i k y \sqrt{3}/2  )|^2$
in the case of a hexagonal lattice generated by three-wave beam interference [Fig.~\rpict{lattices}(c)]. Other configurations of optical patterns can be used to induce Bessel\cite{Kartashov:2004-93904:PRL,Kartashov:2004-444:JOB,Kartashov:2004-65602:PRE,Kartashov:2005-43902:PRL,Kartashov:2005-637:OL,Xu:2005-1774:OE,Xu:2005-1180:OL,Kartashov:2005-1366:JOSB,Kartashov:2005-10703:OE,Wang:2006-83904:PRL,Wang:2006-7362:OE,Fischer:2006-2825:OE}, Mathieu\cite{Kartashov:2006-238:OL}, and quasi-periodic\cite{Freedman:2006-1166:NAT} lattices.

\pict[0.9]{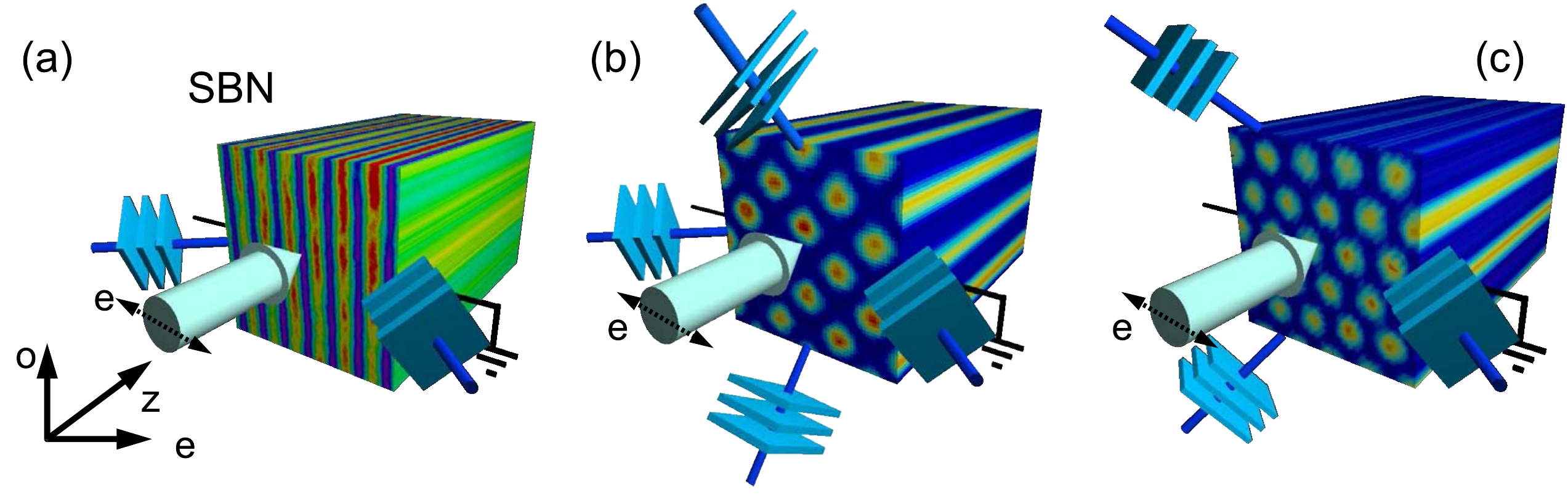}{lattices}{Optical induction of (a) one- and (b,c) two-dimensional photonic lattice in a biased photorefractive SBN crystal of square and hexagonal lattice, respectively.}

The advantage of using optically induced lattices as a test-bed physical system comes from the fact that the inter-site coupling can be easily controlled via the applied bias voltage $V_0$ (positive or negative) or lattice-forming beam intensity, while the Bragg scattering can be controlled via the period of the lattice, which is inversely proportional to the angle between the interfering beams. A disadvantage of the system is the relatively low index modulation, typically of the order of few times $10^{-4}$. This disadvantage results in the need of a rather large lattice period, and relatively short propagation distances (typical crystal is 1-2\,cm long). The possibility for dynamic reconfiguration of the lattice, however, makes the optically induced lattices an attractive tool for studies of nonlinear phenomena in periodic photonic structures of various geometries. Indeed, this technique was applied by several groups and the experimental demonstration of various fundamental effects came quickly\cite{Fleischer:2003-23902:PRL,Neshev:2003-710:OL,Fleischer:2003-147:NAT,Martin:2004-123902:PRL,Fleischer:2005-1780:OE}.

\subsection{Bandgap structure and tunability}

Even though the technique of optical induction results in weak index modulation, this can be sufficient for the appearance of a distinct bandgap structure for the transverse components of the wave vectors $K_x, K_y$ due to the Bragg scattering of waves propagating at small angles. The propagation of linear waves through a periodic lattice is described by the spatially extended eigenmodes, called Bloch waves. They can be found as solutions of linearized Eq.~\reqt{nls} in the form $E = \psi(x,y) \exp( i \beta z + i K_x x + i K_y y)$, where $\psi(x,y)$ has the periodicity of the underlying lattice, and $\beta$ is the propagation constant.

\pict[0.99]{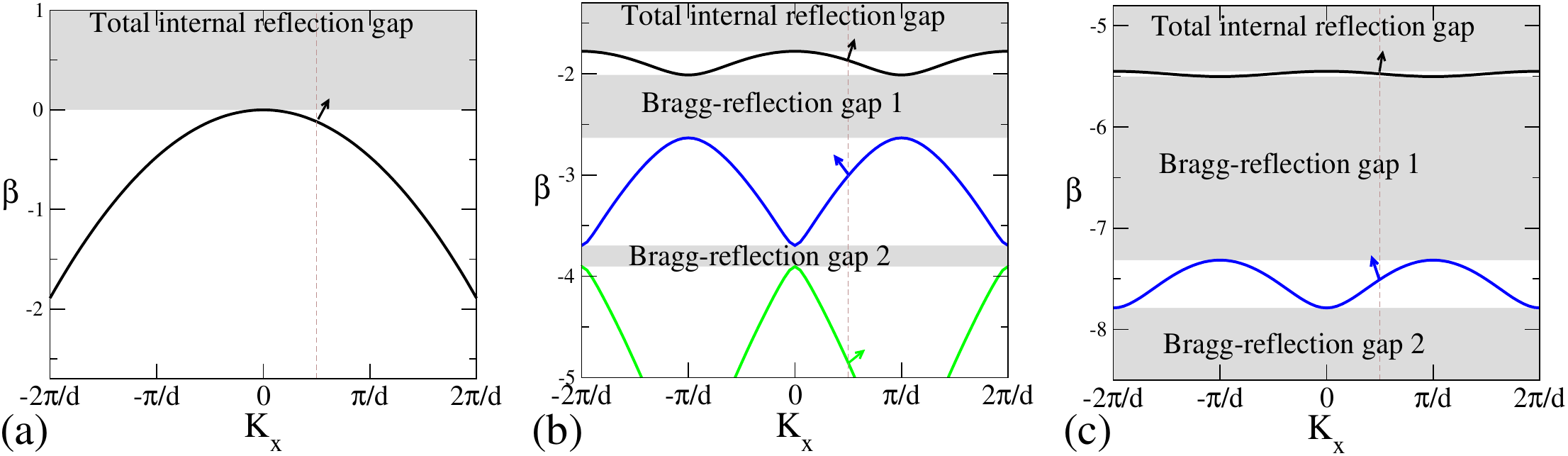}{bandgap}{(a) Dispersion of linear waves in a bulk medium, and (b,c)~Bloch-wave dispersion in (b)~weaker and (c)~stronger 1D optical lattices. Arrows indicate propagation directions at selected dispersion points $K_x = \pi/(2 d)$. }

When the applied bias voltage is zero, the photorefractive crystal is homogeneous and the dispersion relation for the propagation of waves is parabolic, $\beta = - D (K_x^2 + K_y^2)$, as shown in Fig.~\rpict{bandgap}(a). Once a voltage is applied across the crystal, the dispersion is modified, and there appear forbidden gaps in the transmission spectrum\cite{Yeh:1988:OpticalWaves,Russell:1995-585:ConfinedElectrons}, as shown in [Fig.~\rpict{bandgap}(b)] for the case of 1D lattice. Note that the dispersion curves become periodic, $\beta(K_x) = \beta(K_x + 2 \pi / d)$, where $d$ is the lattice period, and accordingly are fully characterized by their values in the first Brillouin zone, $-\pi/d \le K_x \le \pi/d$. The 1D band-gaps open around the points with $K_x d / \pi = 0, \pm 1, \pm 2, \ldots$, where the Bragg scattering condition is satisfied.

The modification of the dispersion relation dramatically changes the beam propagation. The direction of propagation is determined by the normal to the dispersion curves [see the arrows in Figs.~\rpict{bandgap}(b),(c)], while diffraction is determined by the curvature at the corresponding point. Thus waves associated with the top of the first band and propagating along the lattice will experience normal diffraction, while waves corresponding to the bottom of the first band (convex curvature) will experience anomalous diffraction\cite{Pertsch:2002-93901:PRL}. The curvature of the dispersion curves changes in between those two zones from concave to convex, therefore there will be a point where the curvature is zero. Waves belonging to this zone will, therefore, experience no second-order diffraction, similar to the self-collimation effect in photonic crystals\cite{Rakich:2006-93:NAMT}.

It is important to note that the size of the forbidden gaps is proportional to the induced index modulation. An increase of the index modulation leads to flattening of the bands and therefore a wider Bragg reflection gap. Therefore, if one increases the index modulation by increasing the applied bias field, one can dynamically control the direction of propagation of waves inside the structure. As illustrated in Figs.~\rpict{bandgap}(b) and~(c), the increase of the applied voltages makes the waves with a normalized transverse momentum of 0.25 to propagate under smaller and smaller angles inside the structure. Such dynamic tuning of wave propagation is not an isolated phenomenon for optically-induced lattices, but can be applied to other types of periodic structures with electro-optic tunability. In particular, various possibilities for beam control were demonstrated in liquid-crystal cells with periodical modulation induced by an array of electrodes\cite{Fratalocchi:2005-174:OL,Fratalocchi:2005-51112:APL,Fratalocchi:2005-1808:OE,Fratalocchi:2005-51112:APL,Brzdakiewicz:2005-107:OER,Fratalocchi:2005-66608:PRE,Fratalocchi:2006-191:MCLC}.

\pict[0.9]{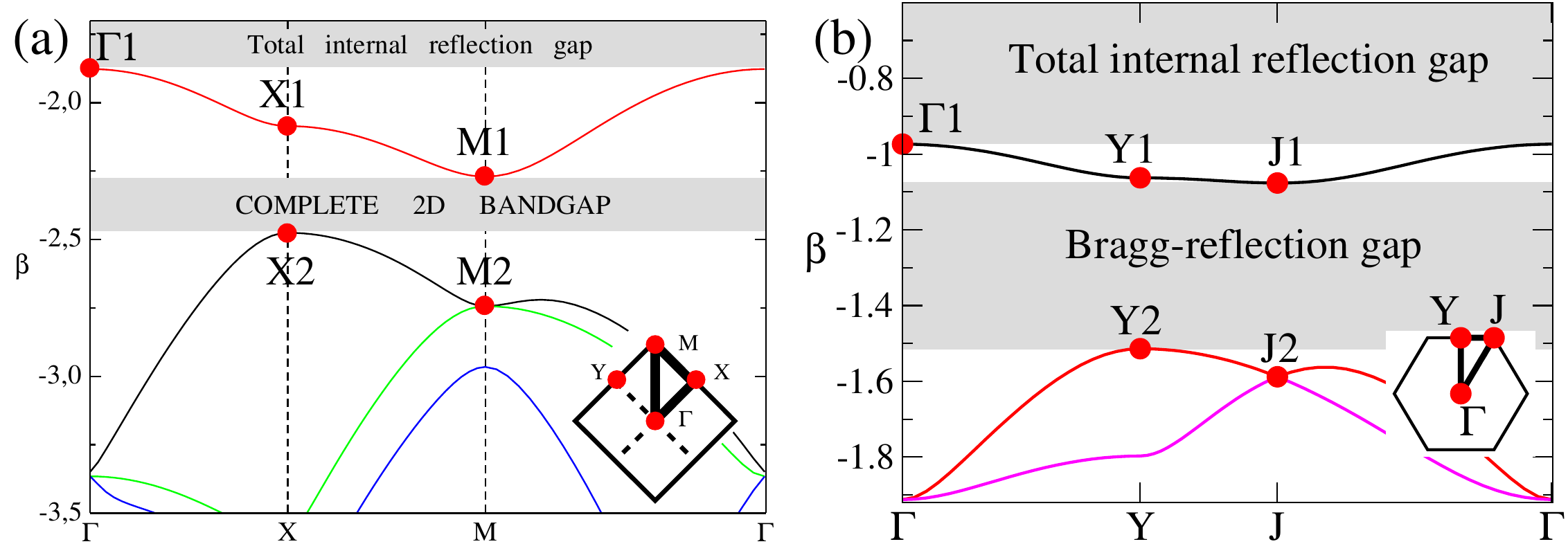}{bandgap2D}{Bloch-wave dispersion in 2D optical lattices of (a)~square and (b)~triangular symmetries. The horizontal axes correspond to the contours passing through the high-symmetry points of the Brillouin zones shown in the inset.}

The optically induced bandgap structure is also well defined in the case of periodic lattices with higher dimensionality. In 2D lattices, the band-gaps may appear only above the critical value of the lattice depth. Examples of characteristic bandgap structures for square and triangular lattices are presented in Fig.~\rpict{bandgap2D}. Note that a complete 2D bandgap appears between the first and the second band. This fact underlines the importance of the optically-induced lattices as an {\em analogue of 1D and 2D photonic crystals}. Therefore, the optically induced lattices represent an accessible test-bed for studies of generic bandgap phenomena in photonic periodic structures.

\subsection{Selective excitation of Bloch waves}

In order to explore the ability for control and steering of beams in an optical lattice, it is important to understand the character and the profiles of the propagating linear waves. Finite beams propagating inside the lattice can be naturally represented as a superposition of Bloch waves of the induced periodic potential. The Bloch wave profiles corresponding to the top and bottom of the first band and the top of the second band of the 1D lattice bandgap spectrum [Fig.~\rpict{Bloch_waves}(a)] are given in Fig.~\rpict{Bloch_waves}(b, solid line), together with the underlying periodic index modulation Fig.~\rpict{Bloch_waves}(b, shading). These Bloch waves have the periodicity of the lattice itself, but can also have a nontrivial phase structure. The Bloch wave from the top of the first band is periodic, but always sign-definite with a flat phase front. On the other hand, the Bloch waves from the bottom of the first band and the top of the second band change sign at each lattice period. Furthermore, the maxima of the Bloch wave intensity associated with the bottom of the first band are centered on the maxima of the refractive index modulation, whereas the intensity maxima of the Bloch waves corresponding to the top of the second band are centered in between the lattice maxima.

\pict[0.9]{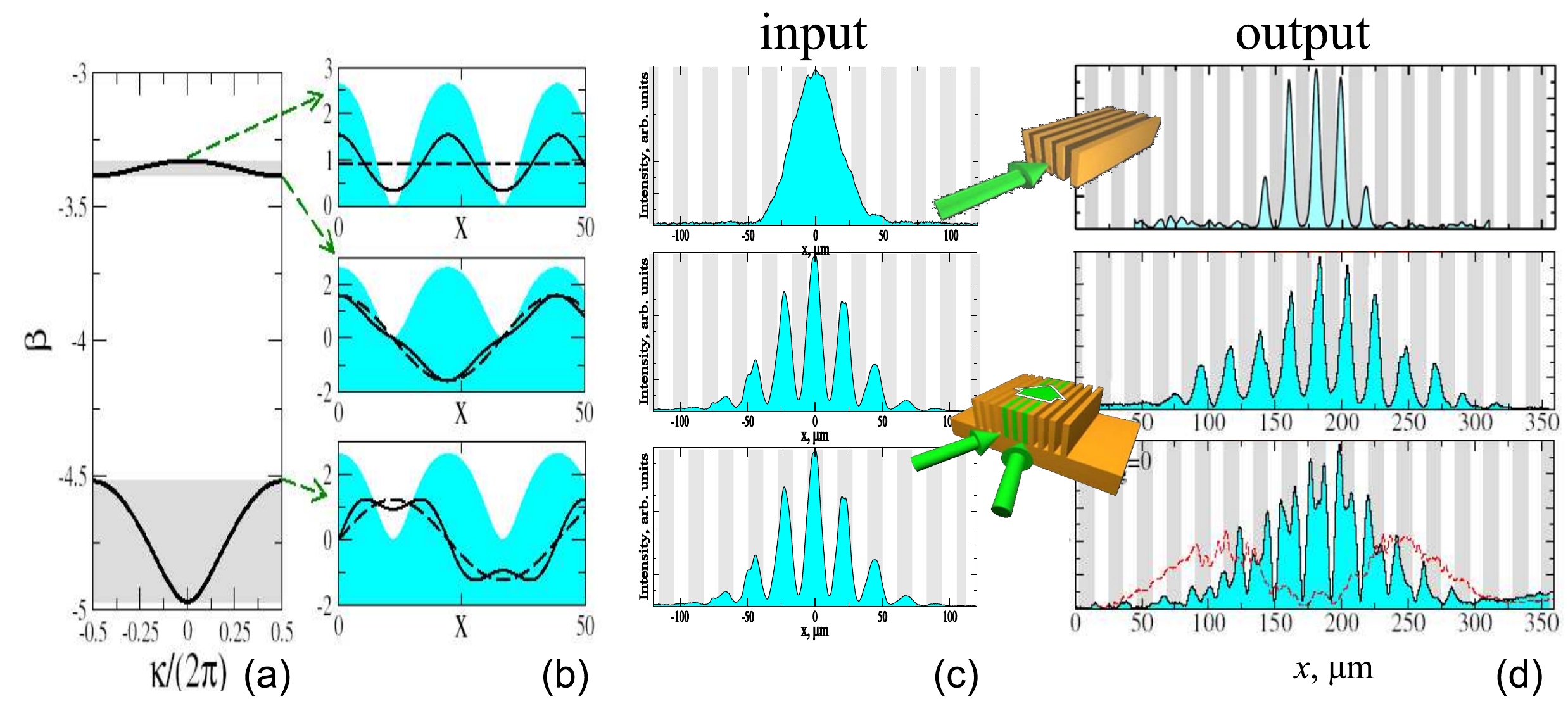}{Bloch_waves}{Bloch waves and their excitation: (a) typical bandgap diagram for 1D optical lattice. (b) Profiles of the Bloch waves (solid line) corresponding to the edges of the first and second bands, superimposed on the leading-order Fourier component (dashed line) and induced refractive index change (shading). (c,d) Experimental excitation of Bloch waves: input and output of the lattice, respectively. Insets show the excitation geometry. Dashed line - output beam profile without the lattice.}

Each Bloch wave has a well defined propagation direction and diffraction coefficient, following the dispersion curves of the bandgap diagram. In order to explore their specific properties, it is important to know how to selectively excite each Bloch wave. Experimentally, this means that most of the light entering the lattice should be efficiently coupled to the desired Bloch wave. For best excitation one needs to exactly match the field profile of the corresponding Bloch wave. Exact matching of the Bloch wave profile, however, is not always possible. A good approximation is given by the leading-order Fourier component of the Bloch wave profile, plotted in Fig.~\rpict{Bloch_waves}(b) with a dashed line. The leading-order Fourier component for the Bloch wave associated with the top of the first band is a constant beam intensity, while for the Bloch waves from the edges of the Bragg reflection gap, the leading order Fourier component is cosine or sine functions. Having this in mind, we demonstrated experimentally efficient excitation of the Bloch waves by a broad Gaussian beam for the Bloch wave of the top of the first band and by two-beam interference for the Bloch waves from the bottom of the first band and top of the second band, respectively\cite{Rosberg:2005-5369:OE}. The only difference in the excitation between the latter two is the position of the interference pattern with respect to the lattice [see Fig.~\rpict{Bloch_waves}(c)], being centered at or in between the lattice sites. The position of the interference maxima is easily controlled by the relative phase between the two input beams (see the bottom inset in Fig.~\rpict{Bloch_waves}). The experimental profiles of the excited Bloch waves at the crystal output are shown in Fig.~\rpict{Bloch_waves}(d) and clearly demonstrate that our technique leads to efficient excitation of the corresponding Bloch waves. This is particularly evident in the excitation of the Bloch waves from the top of the second band which have a characteristic double peak structure, successfully reproduced in our experiments. Other approaches for selective Bloch-wave excitation were also demonstrated based on wave-vector matching with side-on beam incidence\cite{Mandelik:2003-53902:PRL} or prism coupling\cite{Ruter:2006-2768:OL}. Note that if the input is not matched to a particular mode, then a spectrum of Bloch waves can be excited simultaneously from several bands, demonstrating complex interactions in the nonlinear regime\cite{Mandelik:2003-253902:PRL,Cohen:2003-113901:PRL,Sukhorukov:2003-113902:PRL,Sukhorukov:2004-93901:PRL,Buljan:2004-223901:PRL,Bartal:2005-163902:PRL,Cohen:2005-500:NAT,Motzek:2005-2916:OE,Pezer:2005-5013:OE,Bartal:2006-483:OL,Manela:2006-2320:OL,Bartal:2006-73906:PRL}.

\pict[0.9]{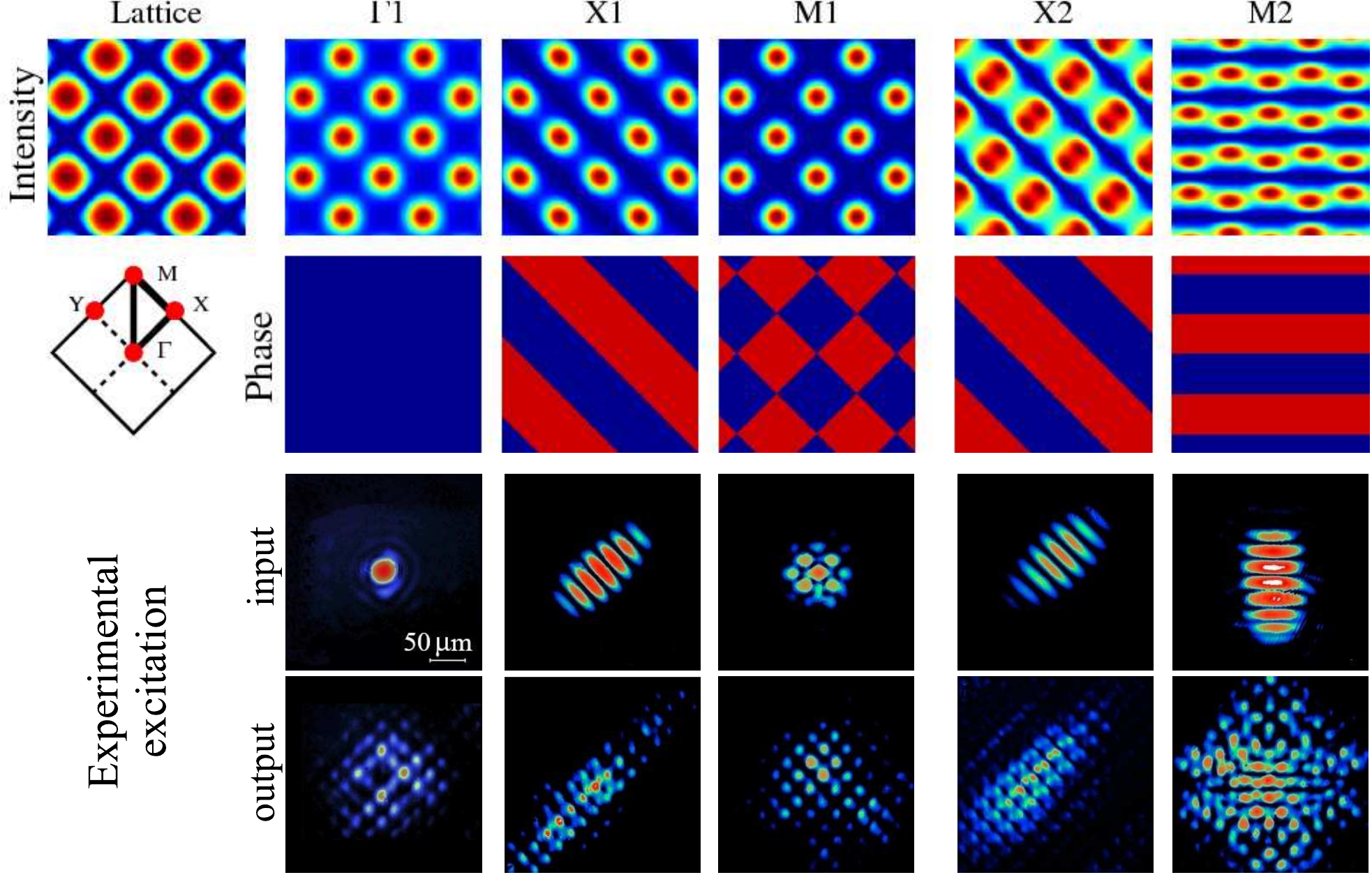}{2D_Bloch_waves}{
2D Bloch waves and their excitation: Intensity (top row) and phase (second row) of different Bloch modes from the high symmetry points of the first and second bands of a square lattice. The blue color for the phase distribution corresponds to the zero phase, while the red color corresponds to the $\pi$ phase. Bottom rows: experimental intensity profiles for excitation of 2D Bloch modes at the input (third row) and the output (bottom row) of the lattice.
}

We have also realized selective Bloch-wave excitation in 2D lattices of square\cite{Fischer:2006-23905:PRL,Trager:2006-1913:OE} and triangular\cite{Rosberg:2007-397:OL} symmetries. The calculated Bloch-wave intensity and phase structure for the high symmetry points of the 2D square lattice from the first and second spectral bands are shown in Fig.~\rpict{2D_Bloch_waves}. The upper row shows the Bloch-wave intensity profiles and the next row shows the corresponding phase structure. As a reference, the first column shows the light intensity of the lattice itself and the corresponding Brillouin zone. For the two-dimensional Bloch waves from the first band, the intensity distribution of all modes reflects the structure of the square lattice, with the intensity maxima coinciding with those of the lattice. However, their phase structures differ substantially. As can be seen from Fig.~\rpict{2D_Bloch_waves}, the phase of the two-dimensional Bloch waves originating from the $\Gamma_1$ point is constant. The phase structure becomes nontrivial for the modes from the X$_1$ and M$_1$ points. For the X$_1$ (Y$_1$) point, the phase represents a stripe-like pattern being constant along one principal direction of the lattice and exhibiting $\pi$ phase jumps along the other direction. For the Bloch waves originating from the M$_1$ point, the phase distribution resembles a chessboard pattern. On the other hand, the two-dimensional Bloch modes from the second spectral band have the intensity maxima centered between the maxima of the square lattice. The phase structure has a form of stripes oriented along one of the principal directions of the two-dimensional lattice for the X$_2$ point, or in $45^\circ$ with respect to the principal axes for the M$_2$ point.

The difference in the phase structure of the two-dimensional Bloch waves translates into differences in the propagation dynamics for beams of a finite size which spectrum is localized in the vicinity of the corresponding high-symmetry points in the Brillouin zone. Indeed, the alternating phase is a signature of strong Bragg scattering, that may lead to an enhanced diffraction of beams similar to the effect of the dispersion enhancement in the Bragg gratings\cite{Slusher:2003:NonlinearPhotonic}. Therefore, the beams can experience anisotropic diffraction due to the asymmetric phase structure of the corresponding Bloch waves, and this can be detected by analyzing the beam broadening in the linear regime.

In order to study experimentally the generation, formation, and propagation of Bloch waves in two-dimensional photonic lattices, we have developed a special setup to match the structure of the input optical beam and the selected Bloch wave, see the third row in Fig.~\rpict{2D_Bloch_waves}. This is achieved by the use of the programmable phase modulator that converts the initially Gaussian probe beam into the desired amplitude and phase modulation at the front face of the photorefractive crystal. A finite beam will diffracts with a rate depending on the value of the diffraction coefficients along the principal directions of the lattice. This allows one to characterize the lattice dispersion by analyzing the diffracted beam patterns shown in Fig.~\rpict{2D_Bloch_waves}, bottom row.

\subsection{Tunable refraction}

The efficient excitation of Bloch waves can be used in experimental schemes for steering and switching of beams in periodic structures. For example, coupling of light into a particular Bloch wave can dramatically change the direction of beam propagation and thus the transport of energy inside the structure can be controlled. To demonstrate this idea, we performed experiments on tunable beam refraction and beam steering in optically induced lattices\cite{Rosberg:2005-2293:OL}. In a straight lattice, all Bloch waves from the top and bottom of each band propagate exactly along the lattice (zero transverse wavevector component). The different Bloch waves, however, can be separated if the lattice is tilted at a small angle as shown in Fig.~\rpict{tunable_refraction}(a). Such a lattice tilt translates into a tilt of the bangap structure, causing Bloch waves from the first band to bend (or refract) along the tilt of the lattice, while Bloch waves from the second band bend in the opposite direction [Fig.~\rpict{tunable_refraction}(a)]. The change in the direction of propagation leads to an output beam shift if the initial light is coupled to a specific Bloch wave.

\pict[0.99]{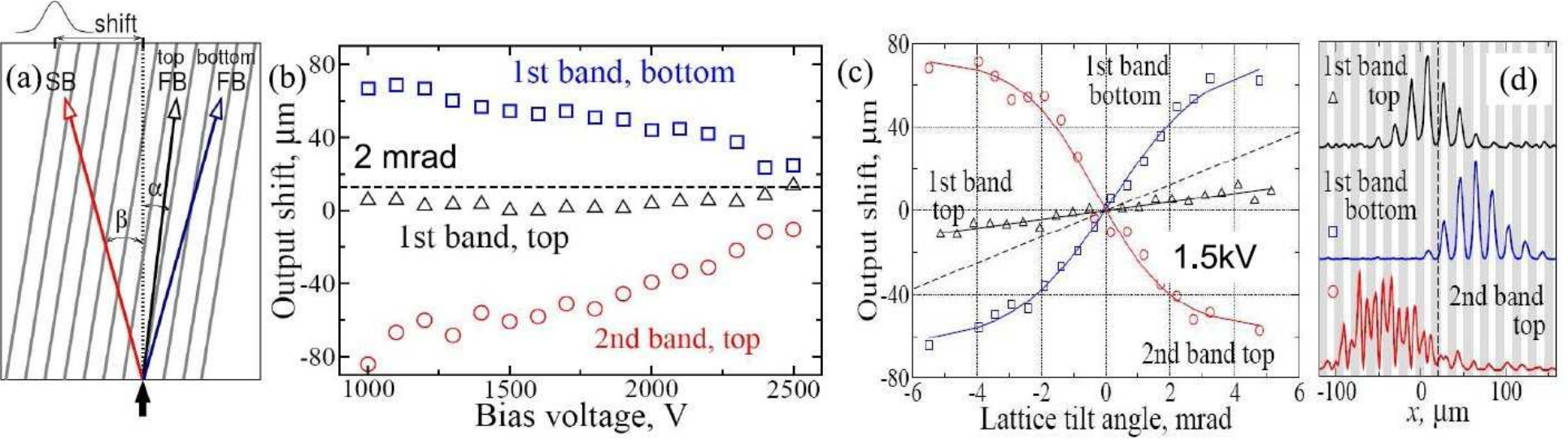}{tunable_refraction}{Tunable refraction in a tilted lattice: (a) Scheme of beam propagation. (b) Tuning of beam refraction for the different Bloch modes by varying the applied bias field. The lattice tilt is 2\,mrad. (c) Tunable refraction by varying the lattice tilt at a constant bias field 3\,kV/cm. (d) Corresponding output intensity profiles for a lattice tilt of 3\,mrad.}

As already discussed, the direction of propagation of each Bloch wave will depend on the depth of the induced refractive index modulation and therefore the output shift can be also tuned by changing the bias voltage applied to the crystal. Experimentally, we measured this shift as a function of the applied voltage across our 5\,mm thick crystal. Our experimental results are depicted in Fig.~\rpict{tunable_refraction}(b). It is clear from the figure that with increasing index modulation (bias voltage $\sim2.5$\,kV), all the Bloch waves approach the tilt of the lattice given by the dashed line in the figure. At lower voltages ($1-2$\,kV), however, we observe that Bloch waves from the top of the first band are shifted less than the lattice, while those from the bottom of the first band and top of the second band are shifted substantially more. We measured a six-fold increase of the lattice shift (at 1\,kV) with positive and negative gain for Bloch waves associated with the bottom of the first band and top of the second band, respectively. These experimental observations demonstrate an important example of tunable refraction in periodic structures, pointing towards potential applications in beam steering technologies.

Furthermore, we measured the dependence of the beam shift at the output vs. the lattice tilt [Fig.~\rpict{tunable_refraction}(c)]. Our results showed that for a small lattice tilt, the dependence is linear, while it saturates to a value equal to the Bragg angle inside the lattice for Bloch waves from the bottom of the first band and top of the second band. The corresponding output beam profiles of the three different Bloch waves are shown in Fig.~\rpict{tunable_refraction}(d). They show that in each case the input light is coupled directly to a particular Bloch wave, with no mixed excitation, and that the beams experience a strong transverse shift.

\section{Nonlinear localization and gap solitons}
\label{sect:nonlinear}

In order to confine optical beams inside the periodic structure it is necessary to balance the beam diffraction through nonlinear self-action. However, as we already know, the diffraction of beams can be significantly modified by the effects of periodicity. Beams associated with the top of each band have normal diffraction which can be balanced by a self-focusing type nonlinearity. Beams from the bottom of each band, on the other hand, exhibit anomalous diffraction, which can be balanced by a self-defocusing type nonlinearity.

\subsection{Solitons in periodic structures}

In a bulk medium with Kerr-type nonlinearity, the propagation of light of high intensity will result in a local refractive index change of the material. For focusing nonlinearity, this index change is equivalent to an induced optical waveguide [Fig.~\rpict{solitons}(a, top)] which can trap and guide the beam in the structure, resulting in the formation of spatial solitons\cite{Stegeman:1999-1518:SCI} [Fig.~\rpict{solitons}(b, top)]. In the case of defocusing nonlinearity, the induced index change is negative [Fig.~\rpict{solitons}(a, bottom)] and will result in anti-guiding of the light, or beam self-defocusing. Therefore, no localized (``bright'') solitons can exist in the case of bulk defocusing material [Fig.~\rpict{solitons}(b, bottom)].
This picture, however, changes fundamentally in periodic structures. In the case of focusing nonlinearity solitons exist near the top of each band, where diffraction is normal. This can result in the formation of discrete spatial solitons from the top of the first band\cite{Christodoulides:1988-794:OL}. Discrete solitons can be excited by a narrow input beam launched into a single lattice site [Fig.~\rpict{solitons}(c, top)]. They were first observed in AlGaAs waveguide arrays\cite{Eisenberg:1998-3383:PRL,Morandotti:2001-3296:PRL} and later reproduced in optically induced lattices\cite{Fleischer:2003-23902:PRL,Neshev:2003-710:OL}, and liquid crystal waveguide arrays\cite{Fratalocchi:2004-1530:OL}. In the case of higher order bands, spatial solitons have been predicted to exist in the Bragg reflection gaps\cite{Feng:1993-1302:OL,Nabiev:1993-1612:OL}. Their excitation is less trivial, and both side-on excitation\cite{Mandelik:2003-53902:PRL} and head-on excitation with periodically modulated input beams\cite{Mandelik:2004-93904:PRL,Neshev:2004-83905:PRL} have been demonstrated [Fig.~\rpict{solitons}(d, top)]. The latter appears beneficial when the transverse soliton velocity needs to be controlled, including the special case of excitation of immobile gap solitons.

\pict[0.9]{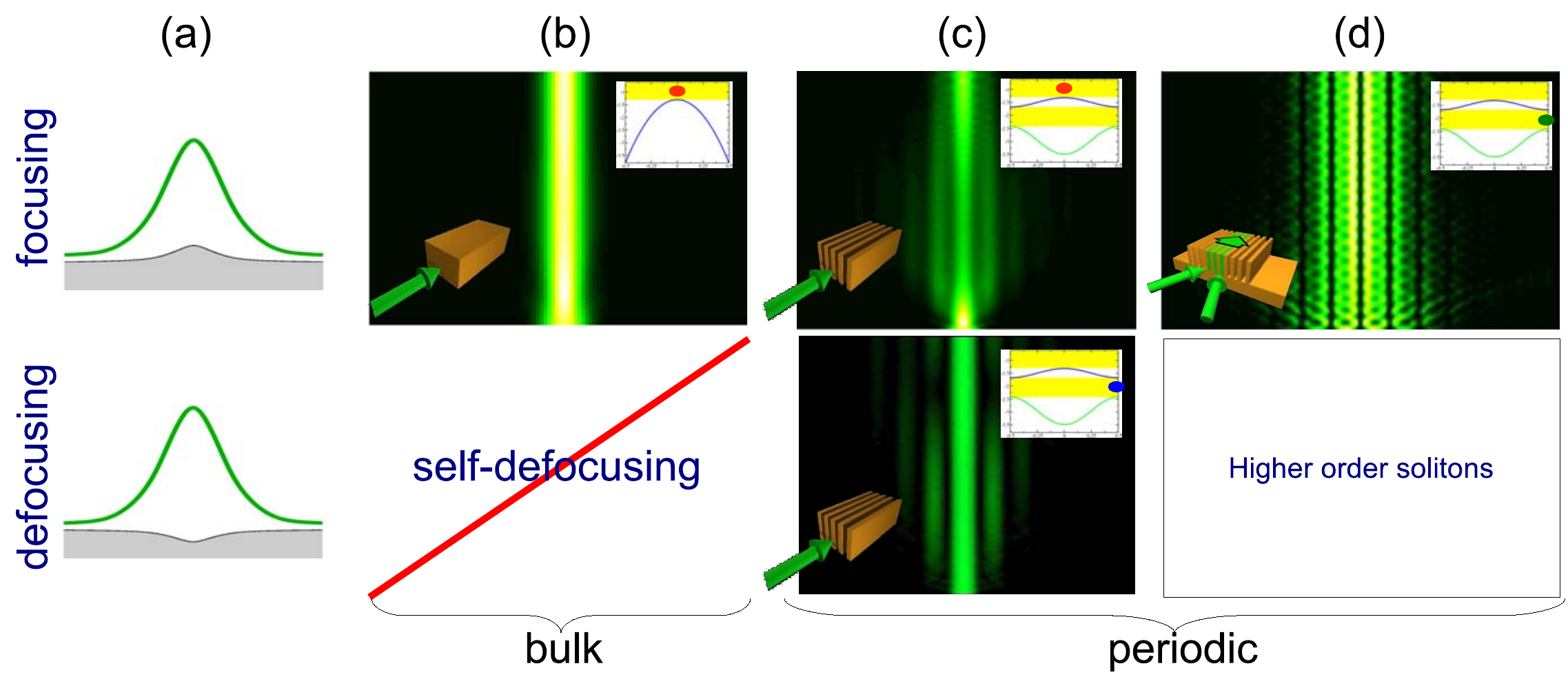}{solitons}{Formation of spatial solitons: (a) self-focusing and self-defocusing nonlinearity will induce a positive or negative defect in the material, respectively. (b) In a bulk material solitons can form only for self-focusing nonlinearity. In a periodic medium solitons can form from the first (c) and second bands (d) for both focusing and defocusing nonlinearity.}

In the case of defocusing nonlinearity 1D solitons exist near the bottom of each band and their propagation constant moves down into the Bragg reflection gap with increasing beam intensity. Solitons in periodic waveguide arrays with defocusing nonlinearity were already predicted in 1993\cite{Kivshar:1993-1147:OL}. They possess a staggered phase structure and the propagation constant lies inside the first Bragg reflection gap [Fig.~\rpict{solitons}\,(c, bottom)]. Therefore, such solitons have properties similar to both discrete and gap solitons in self-focusing nonlinear materials. Herewith we will refer to them as {\em staggered solitons}. Staggered solitons were first demonstrated experimentally by excitation with an inclined input beam\cite{Fleischer:2003-23902:PRL,Iwanow:2004-113902:PRL}. With this method, only a fraction of the input light is coupled into the staggered soliton mode. Later, the generation of staggered solitons was demonstrated in defocusing lithium niobate waveguide arrays by excitation with a TEM$_{10}$ laser mode\cite{Chen:2005-4314:OE,Shandarov:2005-897:TPL}. Such excitation leads to coupling to the unstable ``even'' soliton mode, which then transforms to the stable ``odd''-symmetry mode. The similarity of the staggered solitons and discrete solitons, however, points to a simpler excitation method of high efficiency. As we recently demonstrated\cite{Matuszewski:2006-254:OE} efficient excitation of staggered solitons can be achieved with a narrow input beam coupled to a single lattice site [see Fig.~\rpict{solitons}(c, bottom)], provided the refractive index contrast of the structure is high enough. In this case the beam propagation inside the lattice is very similar to the propagation in lattices with focusing nonlinearity, with the only difference being the staggered phase structure of the output beam. Therefore, one can see that if the optical lattice can be treated as a discrete system, there is a direct analogy between the localization of light for focusing and defocusing nonlinearities.

\subsection{Two-dimensional solitons and their mobility}

The localization of optical beams is even more interesting and nontrivial in the case of higher dimensions. In a 2D lattice, for example, waves can experience anisotropic diffraction\cite{Hudock:2004-268:OL}, which strongly affects the formation of solitons. For a square lattice with a period of 23\,$\mu$m [Fig.~\rpict{reduced-symmetry}(a)] optically induced in a photorefractive SBN crystal, the typical bandgap structure folded along a contour through the high-symmetry points of the lattice is shown in Fig.~\rpict{bandgap2D}(a). For self-focusing nonlinearity, solitons are associated with the top of each band, where diffraction is normal. Due to the crossing of higher order bands, there are only two complete gaps in our 2D lattice: the total internal reflection gap and one Bragg reflection gap. This is in sharp contrast to the 1D case, where multiple Bragg reflection gaps usually exist. Therefore, the soliton family in the 2D lattice is restricted to those two gaps. Discrete solitons are associated with the $\Gamma$ point of the first band\cite{Efremidis:2002-46602:PRE,Efremidis:2003-213906:PRL}. They have been observed in the context of optically-induced lattices by several groups\cite{Fleischer:2003-147:NAT,Martin:2004-123902:PRL,Trager:2006-1913:OE}. Staggered solitons have also been observed in the case of defocusing nonlinearity\cite{Fleischer:2003-147:NAT} and exist near the M-symmetry point of the first band.

\pict[0.8]{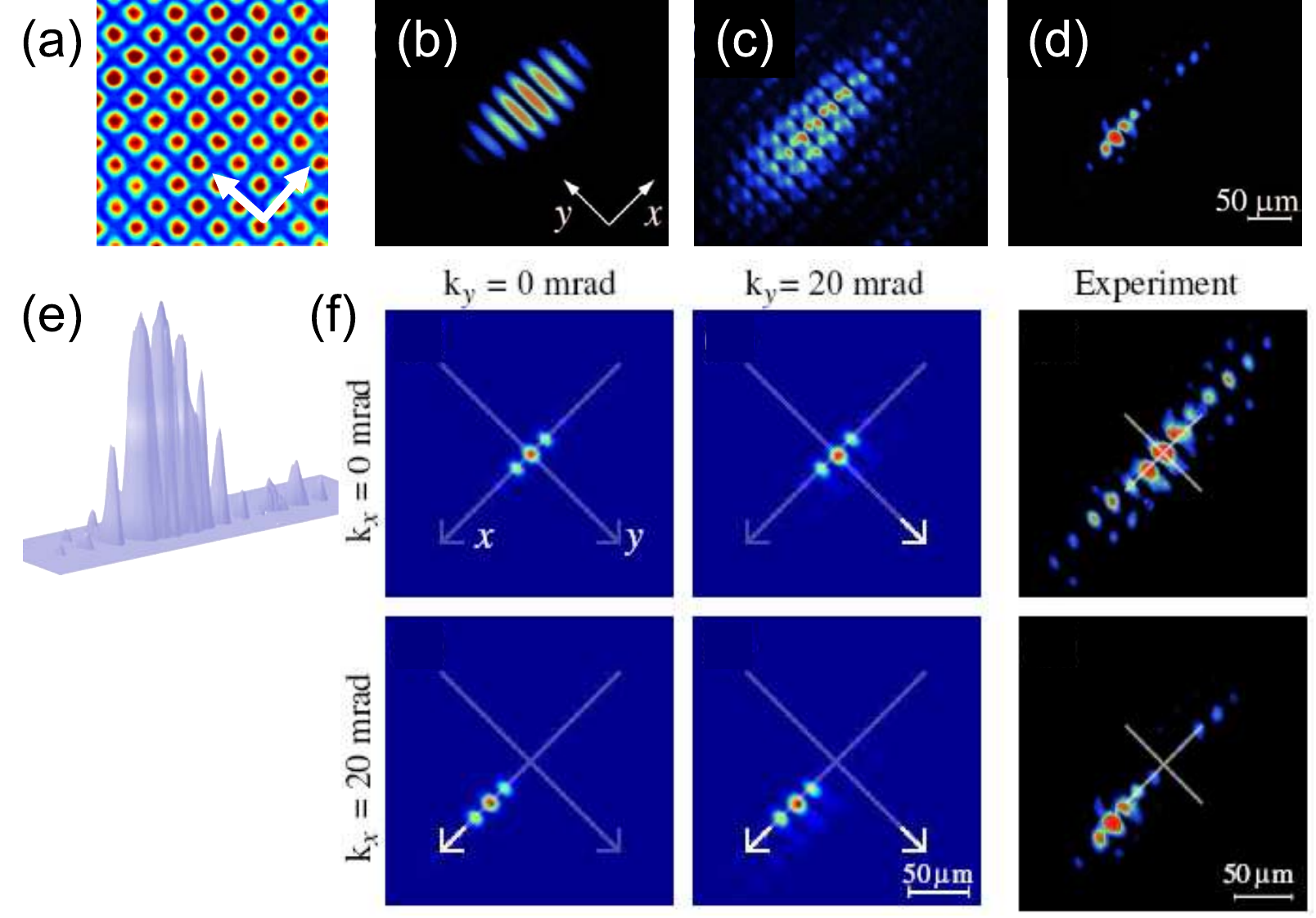}{reduced-symmetry}{Reduced symmetry gap solitons: In a two-dimensional square lattice (a) gap solitons can exist near the top of the second band (X-symmetry point). (b-e) Experimental excitation of reduced symmetry gap solitons: (b) input beam profile; (c) linear diffraction at the output; (d) soliton formation; (e) 3D representation. (f) Anisotropic soliton mobility when initial momentum is applied to the input beams. Left - numerical simulations; right - experimental results.}

Gap solitons in self-focusing nonlinear media have been predicted to exist near the X-symmetry point at the top of the second band\cite{John:1993-1168:PRL,Akozbek:1998-2287:PRE}, having highly anisotropic structure and properties. These solitons are localized in one direction due to Bragg reflection and in the other due to total internal reflection. Their internal structure is fundamentally different from vortex solitons localized in the same gap but having symmetric profiles\cite{Manela:2004-2049:OL,Bartal:2005-53904:PRL}, or discrete solitons which can have elliptical profiles only due to their motion through the lattice\cite{Hudock:2004-268:OL}. To study the X-point gap solitons experimentally\cite{Fischer:2006-23905:PRL} we created a 2D square lattice and used a modulated extraordinarily polarized input beam [Fig.~\rpict{reduced-symmetry}(b)] to match the profile of the Bloch wave associated with the X-symmetry point of the second band. At low laser power the beam diffracts and evolves into the corresponding Bloch wave [Fig.~\rpict{reduced-symmetry}(c)]. Once the input laser power is increased, the output beam experiences a quasi-collapse to a reduced symmetry gap soliton as shown in Fig.~\rpict{reduced-symmetry}(d,e).

The important characteristic of this type of solitons is that they have highly anisotropic mobility properties. The solitons are mobile along their modulated $x$ direction and highly immobile along the $y$ direction. This mobility is illustrated when initial momentum is applied to the soliton. Our numerical simulations and experimental results are summarized in Fig.~\rpict{reduced-symmetry}(f). An initial momentum along the $y$ direction does not lead to any shift of the soliton, but only to a small modulation of its profile, similar to trapping of discrete solitons due to self-induced Peierls-Nabarro potential\cite{Kivshar:1993-3077:PRE,Aceves:1996-1172:PRE,Bang:1996-1105:OL,Morandotti:1999-2726:PRL,Hadzievski:2004-33901:PRL,Maluckov:2005-539:EPB,Meier:2005-1027:OL,Meier:2005-3174:OL,Vicencio:2006-46602:PRE}.
A momentum along $x$, on the other hand, leads to a shift of two lattice sites at the output. Our experimental observations well reproduced the numerical simulations [Fig.~\rpict{reduced-symmetry}(f, right)]. The observed mobility properties of the reduced symmetry gap solitons relate to unique nonlinear transport of beams across the lattice, where the direction of beam propagation is determined by the localized state itself, and not by externally fabricated defects in the structure. The ability to robustly move along one particular direction of the lattice makes the reduced symmetry solitons good candidates for flexible soliton networks in 2D periodic structures. This offers new opportunities compared to the soliton networks suggested earlier for optical signal routing and switching\cite{Christodoulides:2001-233901:PRL}.

Furthermore, the reduced symmetry gap solitons can be regarded as an analogue of a nonlinearity induced waveguide in periodic structures (for mobile solitons)\cite{Fischer:2006-23905:PRL}, or as an optically induced high-Q cavity (for immobile solitons)\cite{Rosberg:2007-397:OL}. We believe that these ideas can be applied in other types of periodic structures such as photonic crystals and microstructured optical fibers.

\section{Fabricated periodic photonic structures}
\label{sect:fabricated}

Even though the optically-induced lattices offer great flexibility for dynamic modification of the lattice parameters, experiments are complex and require extensive stability of the experimental setup in order to avoid fluctuation of the interference pattern of the lattice forming beams and averaging of the lattice potential. A typical experimental setup covers roughly half an optical table, hence the direct applicability of the obtained results is limited. Therefore, it is important to translate the gained knowledge of nonlinear lattices into some more compact fabricated structures, offering closer connection to practical applications and devices.

\pict[0.75]{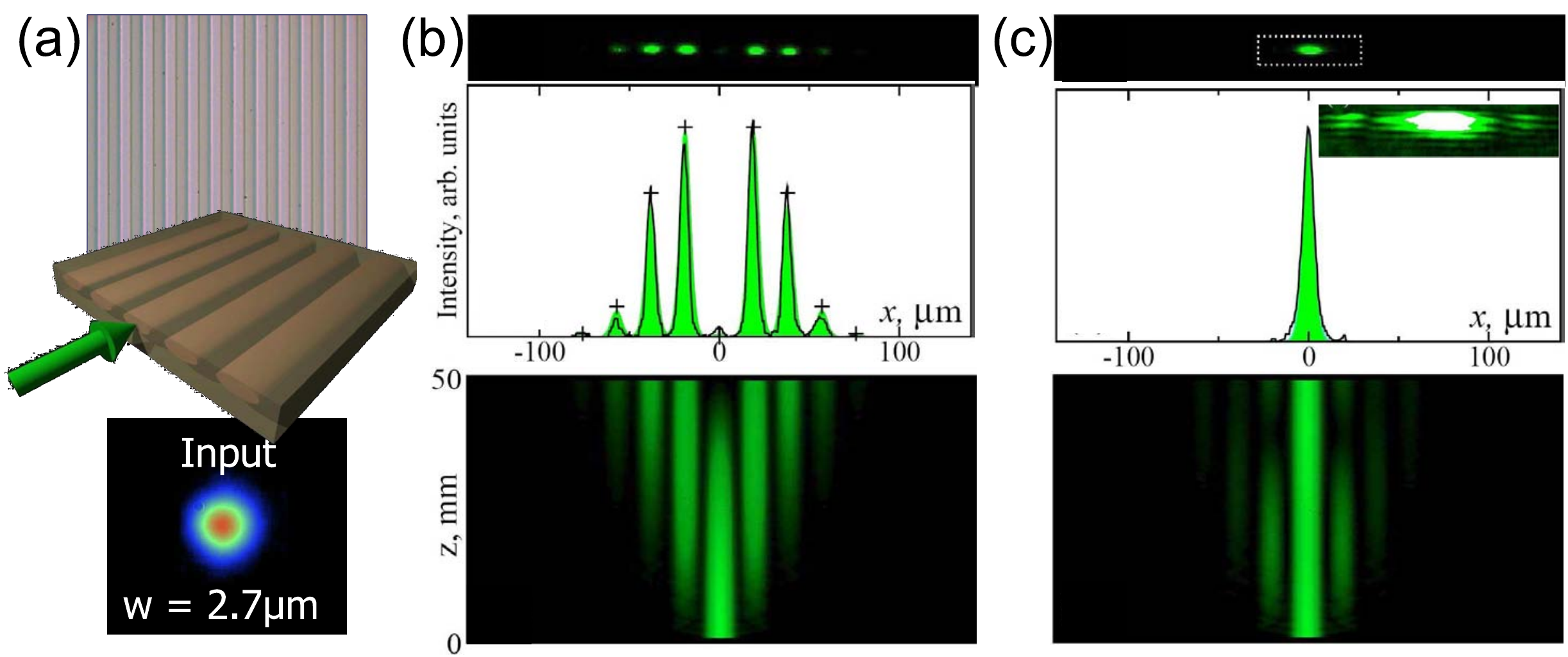}{staggered_solitons}{(a) Fabricated lithium niobate waveguide array, scheme of excitation, and intensity profile of the input beam. (b) Discrete diffraction at low laser power (10\,nW): (top) 2D intensity profile, (middle) transverse intensity distribution and comparison with discrete model (crosses); (bottom) numerically calculated propagation inside the sample. (c) Formation of a staggered soliton at 1\,mW laser power, soliton intensity profile and interferogram revealing the staggered phase structure. (bottom) Numerically calculated nonlinear propagation.}

\subsection{Discrete solitons in defocusing waveguide arrays}

To explore the propagation of light in fabricated periodic photonic structures, we utilized an array of optical waveguides fabricated by titanium indiffusion in a mono-crystal x-cut lithium niobate wafer [see Fig.~\rpict{staggered_solitons}(a, top)]. The array is 5\,cm long and consists of 100 closely spaced waveguides with a period of 19 or 10\,$\mu$m. These fabricated structures offer great simplification of experiments. Furthermore, lithium niobate is known to exhibit strong photovoltaic nonlinearity of defocusing type\cite{Valley:1994-4457:PRA} at micro-Watt power levels for visible wavelengths. To test the linear propagation of light and demonstrate nonlinear localization in such arrays\cite{Matuszewski:2006-254:OE}, we coupled light into a single channel of the array as shown in Fig.~\rpict{staggered_solitons}(a) and monitored the intensity profile at the output with increasing laser power. The input beam was circular with diameter of 2.7\,$\mu$m. At low laser powers (10\,nW) the beam experiences discrete diffraction, where most of its energy is coupled away from the central waveguide [Fig.~\rpict{staggered_solitons}(b)]. This behaviour is well reproduced by calculations using both a discrete model for weakly coupled lattice cites [crosses in Fig.~\rpict{staggered_solitons}(b, middle)], and a more general continuous model corresponding to our array (bottom). When the input laser power is increased light is coupled back into the central guide of the array, and at powers of about 1\,mW all light is concentrated in the central guide of the array [Fig.~\rpict{staggered_solitons}(c)]. An interferometric measurement of the output beam profile reveals that light in the central and the neighboring channels is out of phase, which is a direct indication that the observed localized state has a staggered phase structure and resembles a staggered soliton\cite{Kivshar:1993-1147:OL}.

The power dependent transition between discrete diffraction at low power where almost no light is transmitted inside the central channel, and the single channel localization proves to be beneficial for applications of laser mode locking\cite{Proctor:2005-8933:OE}. In such application the spatial dynamics of light is translated to temporal dynamics and short pulse formation in the laser cavity. Furthermore, the waveguide technology is well compatible with fibers and applications of Kerr-lens mode-locking of fiber lasers leading to high contrast pulse generation seems feasible. The staggered phase structure of the localized state in the case of defocusing nonlinearity also appears to be more advantageous comparing to localization in structures with self-focusing nonlinear response\cite{Proctor:2005-8933:OE}. We expect that the staggered profile can lead to improved contrast of the pulse generation. It is also important to note that the transition from diffraction to nonlinear localization has a well-defined threshold\cite{Meier:2005-1432:JOSB}, in contrast to the localization in homogeneous media. Such a threshold behaviour can be also applied to an optical diode device, replicating the transmission characteristics of usual electronic diodes.

\subsection{Nonlinear surface waves}

Additional functionality for control of light propagation in photonic lattices can be achieved when the optical beam interacts with a defect or an interface (``surface'') of the structure. For example, an optical beam injected at the edge waveguide of the array will be reflected from the interface formed between the periodic and homogeneous parts of the structure [Fig.~\rpict{surface}(a)]. This effective ``repulsion'' from the surface leads to modified discrete diffraction of the beam\cite{Makris:2005-2466:OL} and practically no light remains in the input edge waveguide. At high input powers, however, the surface can support various types of nonlinear surface waves\cite{Makris:2005-2466:OL}. In the case of focusing nonlinear response of the material, such nonlinear surface waves exist in the form of discrete surface solitons\cite{Makris:2005-2466:OL}. They have a propagation constant inside the total internal reflection gap of the periodic structure since the high power light induces a positive index defect at the edge waveguide of the array [Fig.~\rpict{surface}(b, top)]. Such solitons were recently observed experimentally in AlGaAs waveguide arrays\cite{Suntsov:2006-63901:PRL}. In the case of defocusing nonlinearity, surface solitons can also exist\cite{Makris:2005-2466:OL}. Their propagation constant resides inside the Bragg reflection gap and therefore they can be termed as surface gap solitons\cite{Kartashov:2006-73901:PRL}. Due to the defocusing nonlinearity, the high intensity of the light results in a negative index defect at the edge waveguide [see Fig.~\rpict{surface}(b, bottom)]. This defect leads to a less localized nonlinear surface wave which extends deeper inside the bulk medium. This implies that the surface gap solitons can be potential candidates for application of surface sensing and testing. Furthermore, unlike their discrete counterparts in self-focusing media, the surface gap solitons have a well-pronounced staggered phase structure, resulting in zero intensity in between the waveguides.

\pict[0.99]{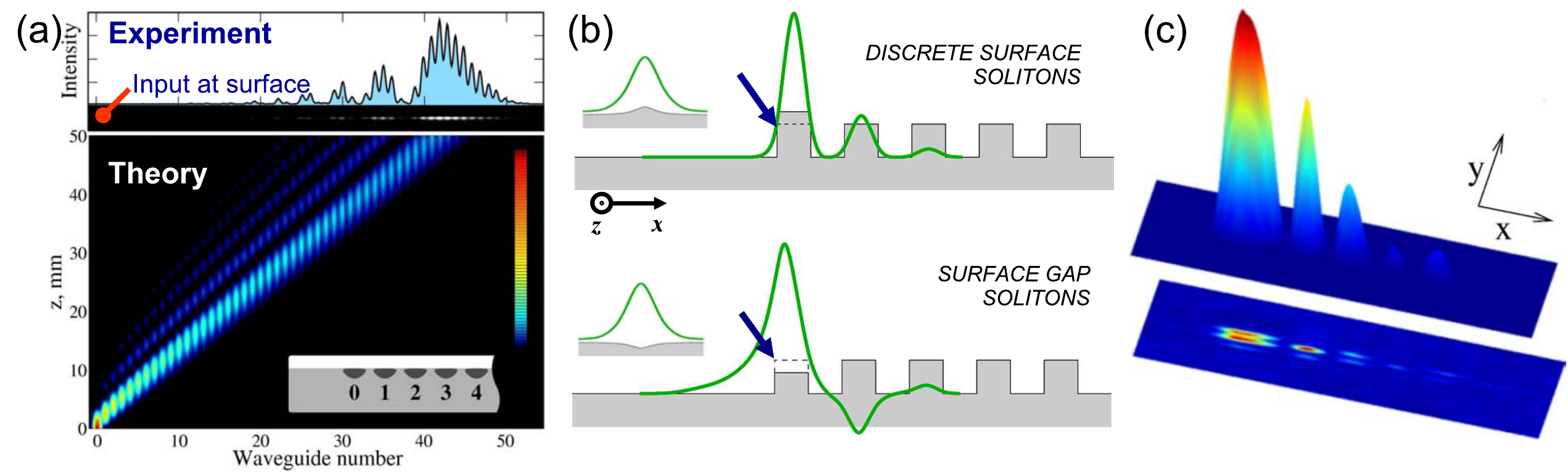}{surface}{Surface waves: (a) Linear repulsion from the surface. (b) Theoretically predicted nonlinear surface waves for focusing (top) and defocusing (bottom) nonlinearities. (c) Experimentally observed surface gap soliton in defocusing lithium niobate waveguide array.}

In our experiments\cite{Rosberg:2006-83901:PRL} we demonstrated the formation of surface gap solitons by injecting a narrow beam at the edge waveguide of a lithium niobate waveguide array. At low laser power (100\,nW) the light is reflected by the surface [Fig.~\rpict{surface}(a, top)], while as the power is increased the beam is attracted to the surface and at a power of 0.5\,mW, a surface gap soliton is formed in the array [Fig.~\rpict{surface}(c, top)]. By use of interferometric measurements we also confirmed the staggered phase structure of the gap soliton. As seen in Fig.~\rpict{surface}(c, bottom) the interference fringes in each neighboring guide are shifted by half a period.

Similar nonlinear surface waves have been demonstrated by other groups, in photorefractive iron-doped lithium niobate waveguide arrays\cite{Smirnov:2006-2338:OL}, and periodically poled lithium niobate arrays with quadratic nonlinearity\cite{Siviloglou:2006-5508:OE}. In the latter case, the nonlinear response is dependent on the coupling between the fundamental beam and the generated second harmonic, and thus the nonlinear response can be tuned from focusing to defocusing by temperature control, adjusting the phase mismatch of the harmonic generation process. Surface solitons may also be supported by optical nonlinearities with a nonlocal response\cite{Kartashov:2006-2595:OL}. More general types of nonlinear localized waves can exist at superlattice boundaries\cite{Molina:2006-2332:OL}, interfaces between different lattices\cite{Makris:2006-2774:OL,Kartashov:2006-2172:OL}, in two-dimensional geometries\cite{Kartashov:2006-2329:OL,Makris:2006-2774:OL,Kartashov:2006-4049:OE}, and at lattice dislocations\cite{Kartashov:2005-243902:PRL,Ablowitz:2006-35601:PRE,Wang:2006-7362:OE}. The presence of interfaces and introduced or dynamically induced defects can also be used to perform tunable beam steering\cite{Krolikowski:1996-876:JOSB,Peschel:1999-1348:APL,Morandotti:2003-834:OL,Sukhorukov:2005-1849:OL,Kartashov:2006-1576:OE,Molina:2006-134:PLA,Smirnov:2006-11248:OE}. Additional possibilities are associated with multiple beam interactions and the formation of vector\cite{Meier:2003-143907:PRL,Hudock:2005-7720:OE} or multi-gap\cite{Cohen:2003-113901:PRL,Sukhorukov:2003-113902:PRL,Garanovich:2006-4780:OE,Kartashov:2006-4808:OE} surface solitons.

\section{Concluding remarks}

In this review, we have described several fundamental effects observed in light propagation in nonlinear periodic structures. We would like to emphasis the ability of periodic structures to control the beam diffraction and refraction by tuning the structural parameters of the photonic lattice. We have also demonstrated how nonlinear self-action of beams can be used to balance their diffraction through the formation of nonlinear localized states, or spatial optical solitons. These solitons can have different properties and characteristics, depending of the type of nonlinearity and their association with the specific Bloch waves of the periodic structure. Furthermore, we have demonstrated how the edges of the periodic structures can affect the nonlinear localization through the formation of nonlinear surface waves.

In addition, we have also pointed towards possible applications of the observed effects. However, it is important to state that a wider range of applications is still to follow. We have shown that most effects are scalable and, therefore, they can be observed in structures with different periodicity. We were able to demonstrate the results in different periodic structures -- from optically-induced lattices to fabricated waveguide arrays. The application of the presented concepts to smaller structures and their integration in photonic devices on an optical chip is an exciting direction to follow. We hope that with the current development of modern technologies and fabrication of photonic crystals a bright future of nonlinear periodic structures for light control on micro and nanoscales is soon to come.

\section*{Acknowledgements}

We would like to thank our numerous collaborators and research students for their substantial and valuable contributions to the results summarized in this review paper. We also acknowledge the support of the Australian Research Council.

\bibliography{abbrev,papers,books,db_art_nonlinear_lattices}

\end{document}